# From Bad to Worse: Using Private Data to Propagate Disinformation on Online Platforms with a Greater Efficiency

Protik Bose Pranto
ppranto@asu.edu
Arizona State University
Tempe, Arizona, USA

Waqar Hassan Khan
wkhan17@asu.edu
Arizona State University
Tempe, Arizona, USA

Sahar Abdelnabi
sahar.abdelnabi@cispa.de
CISPA Helmholtz Center for
Information Security
Saarbrücken, Saarland

Rebecca Weil
weil@cispa.de
CISPA Helmholtz Center for
Information Security
Saarbrücken, Saarland

Mario Fritz
fritz@cispa.de
CISPA Helmholtz Center for
Information Security
Saarbrücken, Saarland

Rakibul Hasan
rhasan3@asu.edu
Arizona State University
Tempe, Arizona, USA

## ABSTRACT

We outline a planned experiment to investigate if personal data (e.g., demographics and behavioral patterns) can be used to selectively expose individuals to disinformation such that an adversary can spread disinformation more efficiently compared to broadcasting the same information to everyone. This mechanism, if effective, will have devastating consequences as modern technologies collect and infer a plethora of private data that can be abused to target with disinformation. We believe this research will inform designing policy and regulation for online platforms.

## 1 PROBLEM SPACE

Technologies collect huge amount of personal and potentially sensitive data, including demographic, bio-metric, behavioral patterns, and (dis)interests. These data may be collected, shared, or used with or without people's consent [1, 5, 11] and digitally stored, making them permanent, replicable, and re-shareable [23], heightening the concern regarding numerous privacy risks, such as data breaches, hacking, identity theft, and other forms of unauthorized access (e.g., [7, 15, 25, 30, 31, 37]). We plan to investigate whether personal information can be abused to propagate disinformation, by exposing individuals to specific disinformation based on their demographic attributes, personality traits, behavioral patterns, physiological and emotional states, or other properties. Thus, our research is at the intersection of privacy and disinformation, where we aim to investigate if private data can be used to "target" individuals with disinformation to elicit desired outcomes, such as increasing the likelihood of believing the information or further propagating it to others, or both.

We hypothesize that an individual, when targeted with disinformation based on some property $p$ (e.g., gender, socio-economic status, interests, personality traits, etc.), will *react differently* compared to other people who do not possess property $p$. For example, health-conscious individuals may react differently to health-related disinformation than individuals with less interest in such news. The reaction can be either positively or negatively correlated with the level of interest (i.e., a higher or lower level of trust in or engagement with the provided information). We aim to investigate the magnitude and direction of change in reaction from individuals targeted with disinformation compared to showing the same information to a random set of people.

Prior research has reported that certain properties, e.g. age, gender, emotional intelligence, etc., are associated with a higher level of trust in fake news. For instance, individuals with low emotional intelligence are more susceptible to false information than individuals with high emotional intelligence [29]. Additionally, older people were found to be more likely to believe false information than younger people [4, 14, 27]. Previous research also found a positive correlation between education level and false news acceptance [8, 27]. Thus, it is likely that by serving different disinformation to different groups of the population that possess these properties, an adversary will be able to propagate disinformation more efficiently than by broadcasting the same message to the whole population.

Moreover, the advancement of Extended Reality, such as Augmented Reality (AR) [6], Virtual Reality (VR) [43] technologies are closing the gap between the physical and virtual worlds [44]. Even in the Metaverse [28], interpersonal engagement is more intense and scalable than in traditional social media environments [38]. Such interactions may allow adversaries to exploit an unprecedented amount of personal information and use them to target people with disinformation.

Conversely, in some cases, targeting may result in a lower level of trust in disinformation. E.g., if someone is targeted based on their interests or hobbies, they might be less likely to believe in that information because of their prior knowledge on that topic of interest [33]. Yet, a vast amount of research suggests that the exact opposite might be true. Due to a processing advantage of familiar information, familiarity with the topic may lead to the impression that the information is true [12, 13]. However, identifying characteristics of people who are resilient to targeted disinformation will benefit future research in crowdsourced fact-checking (e.g., by directing potential disinformation to fact-checkers with these characteristics).

For the cases where targeting leads to a higher level of trust or engagement, the consequences can be disastrous. Since individuals' preexisting ideas and interests are positively correlated with their acceptance of false information [34], targeting may even create opportunities to mislead sub-populations who were previously believed to be resilient to disinformation, e.g., highly educated individuals [8] and young adult [4], if they find the served information aligned with their personal characteristics and interests. For example, young adults who are concerned about health, climate, etc. can be misled by deceptive articles [10, 36], especially if they are emotionally invested in that topic [22]. Furthermore, platform users build networks with others who have similar interests [21] and share information they find interesting with their connections [8, 32], massively scaling up the number of affected people who can further propagate the information at no cost to the adversary.

Additionally, posts generated by Artificial Intelligence (AI) can be highly emotional [9] and can be used to target individuals with low emotional intelligence [29] that can be inferred from breached data [18, 24], sensor data [19], or online personality tests [5]. Moreover, AI is currently used for personalized recommendations on digital platforms, healthcare, and marketing, which depend heavily on user characteristics, behaviors, demographics, and preferences data [16, 20, 35]. Since AI technologies such as DeepFake [39], botnet [2], ChatGPT [40] etc., are cheap, fast, scalable, and able to create personalized content [17], adversaries may soon be capable of generating targeted disinformation automatically and faster than a human being.

## 2 EXPERIMENTAL DESIGN

As a first step towards understanding the effect of targeted disinformation, we designed a study where we target based on people's demographic attributes and topical interests (e.g., healthy diet, cooking, celebrity news, gardening, movie, books, etc.). Each participant will read 20 recently published news articles manually collected from high [26] and low-credible [42] news sources selected depending on NewsGuard's reliability score [41]. Also, the low-credible news sources selected for this study tend to publish false news [3]. However, which news item stems from which source will not be revealed to participants. They will be asked whether or not they believe the news to be true. Suppose we find that people believe false information that matches their self-reported interests more than false information that doesn't match their self-reported interests. In that case, we can infer that targeted disinformation influences people more than any non-targeted misinformation. In particular, if targeted disinformation turns out to be effective in propagating disinformation, this finding will strengthen the case for better protection of consumer data with regulations as well as platform design choices. As such, this study will inform the design of online platforms and policies regarding the collection and use of personal data.